# The Order Parameters for Pairing and Phase Coherence in Cuprates; the Magnetic Origin of the Coherent Gap. The MCS Model of High-$T_c$ Superconductivity

A. Mourachkine

*Université Libre de Bruxelles, Service de Physique des Solides, CP233, Boulevard du Triomphe, B-1050 Brussels, Belgium*

*The phase diagram of order parameters for pairing and phase coherence in hole-doped cuprates is discussed. By examining carefully some recent inelastic neutron scattering data obtained on hole-doped cuprates and heavy fermion compound* $UPd_2Al_3$ *in which the superconductivity is mediated by spin fluctuations, we conclude that the coherent gap in hole-doped cuprates has most likely the magnetic origin and scales with* $T_c$, *on average, as* $2\Delta_c/k_B T_c = 5.4$. *We discuss a model of the superconductivity in hole-doped cuprates and the symmetries of two order parameters.*
  PACS numbers: 74.25.Dw, 74.72.-h, 74.20.Mn

## 1. INTRODUCTION

In general, the superconductivity (SC) requires the formation of the Cooper pairs and the phase coherence among them. In the BCS theory for metals, the mechanisms responsible for the pairing and establishment of the phase coherence are identical - by phonons. Both phenomena occur almost simultaneously at $T_c$. In SC copper-oxides, there is a consensus[1,2] that these two mechanisms occur at different temperatures, at least, in the underdoped regime, $T_{pair}$  $T_c$. The order parameters (OPs) responsible for each process have different dependencies on hole concentration, $p$ in $CuO_2$ planes.[1,3] The magnitude of the OP responsible for phase coherence, $\Delta_c$, which is proportional to $T_c$, has the parabolic dependence[1,4] on $p$. While the magnitude of the OP responsible for pairing, $\Delta_p$ increases linearly with the decrease of hole concentration.[1,3,5] Both the $\Delta_c$ and $\Delta_p$ are SC OPs. The

A. Mourachkine

magnitude of the total energy gap[6] is equal to $\Delta = (\Delta_c^2 + \Delta_p^2)^{1/2}$. However, there is no consensus on the origins of the two OPs. There is an evidence that the spin-exchange interactions play a central role in the cuprates.[7,8] Thus, it is reasonable to assume that one out of the two OPs has the magnetic origin.

In this paper, we discuss the phase diagram and INS data obtained on hole-doped cuprates, and we show that the coherent gap, $\Delta_c$ has most likely the magnetic origin. We discuss also a MCS model of high-$T_c$ superconductivity (HTSC) and the symmetries of the $\Delta_c$ and $\Delta_p$ gaps.

## 2. THE ORDER PARAMETER FOR PAIRING

Figure 1 shows a phase diagram of two energy gaps in hole-doped cuprates.[1] In Fig. 1, the $\Delta_c$ scales with $T_c$ as $2\Delta_c/k_B T_c = 5.45$. The dependence $\Delta_c(p)$ is parabolic since $T_c = T_{c,\,max}[1 - 82.6(p - 0.16)^2]$, where $T_{c,\,max}$ is the maximum $T_c$ for each family of cuprates.[4] There is an opinion that the $\Delta_p$ is a normal-state gap. However, tunneling measurements performed by Miyakawa *et al.* show unambiguously that a tunneling gap with the maximum magnitude in under- and overdoped Bi2212 is a SC gap.[5] In maximum-gap measurements, the Josephson $I_c R_N$ product (both average and maximum), where $I_c$ is the maximum Josephson current and $R_N$ is the normal resistance of a tunnel junction, *increases* with the decrease of hole concentration while the coherent energy range is also decreasing in the underdoped regime (see Fig. 1). The Josephson strength is a characteristic of the coherent state. The $I_c R_N$ product scales with the $\Delta_p$ and not with the $\Delta_c$. Consequently, the $\Delta_p$ is a SC gap.

## 3. THE ORIGIN OF THE COHERENT GAP

Recent inelastic neutron scattering (INS) experiments have shown the presence of sharp magnetic collective mode ('resonance peak') in the SC state of $YBa_2Cu_3O_{6+x}$ (YBCO).[8-13] The discovery of the resonance peak in $Bi_2Sr_2CaCu_2O_{8+x}$ (Bi2212)[14] points out that the resonance peak is an *intrinsic* feature of the SC in the double-layer cuprates studied so far. It is important to note that the resonance-peak position, $E_r$ is in quantitative agreement[15,7] with the condensation energy of YBCO and the temperature-dependent resonance intensity is correlated with the specific heat of YBCO.[8] The resonance peak has been also observed by INS in a heavy fermion compound[16] $UPd_2Al_3$ in which the SC is mediated most likely by spin fluctuations.[17,18] It is also important to note that the SC in $UPd_2Al_3$ coexists

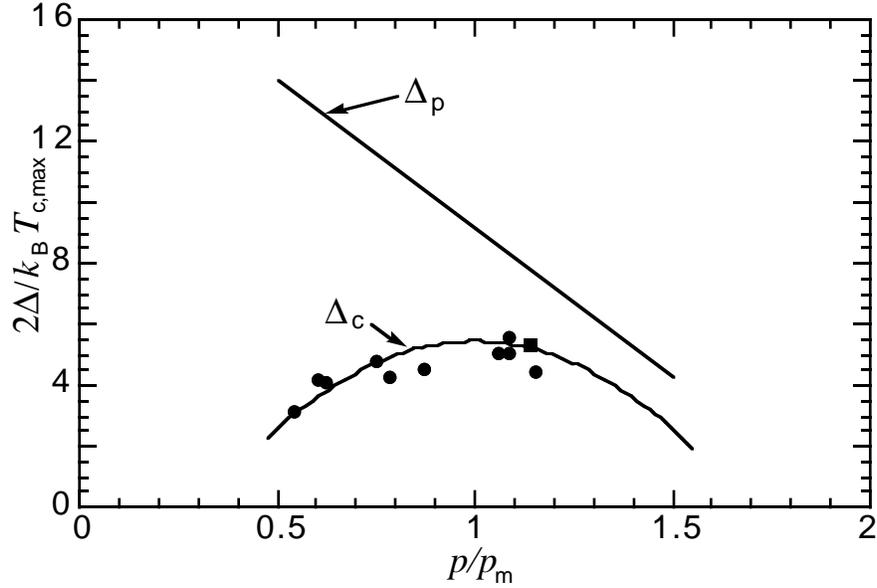

Fig. 1. Phase diagram in hole-doped cuprates: $\Delta_c$ is the coherence energy range, and $\Delta_p$ is the pairing energy gap.[1] INS data: dots (YBCO[8-13]) and square (Bi2212[14]). For more information, see Ref. 19. The $p_m$ is a hole concentration with the maximum $T_c$.

with the antiferromagnetic order like in the cuprates.

In Fig. 1, we present INS data of YBCO[8-13] and Bi2212[14]. One can see in Fig. 1 that there is a good agreement between the $\Delta_c$ and the INS data. Since the resonance peak has been also observed by INS in a heavy fermion compound UPd$_2$Al$_3$ in which the SC is mediated by spin fluctuations it is reasonable to assume that the $\Delta_c$ has most likely the magnetic origin. In general, the SC mediated by spin fluctuations implies that the coherent gap in hole-doped cuprates has the $d_{x^2-y^2}$ (hereafter, d-wave) symmetry.[7] The average $2\Delta_c/k_B T_c$ value for the INS data presented in Fig. 1 is equal to 5.38. More information on this issue can be found elsewhere.[19]

### 3. THE MCS MODEL OF THE SC IN CUPRATES

There are two very good candidates for the model of HTSC[1]: a stripe model[20] and bipolaron model.[21] However, both models can not take account of the magnetic origin of the $\Delta_c$. Recently, we proposed a MCS (Magnetic

Coupling of Stripes) model[22] which is based upon the stripe model[20] which is in turn based upon a spinon SC along charge stripes. The main difference between the two models is that the coherent state of spinon SC is established differently in the two models, by spin fluctuations into antiferromagnetic domains of $CuO_2$ planes in the MCS model, and by the Josephson coupling between stripes in the stripe model. Thus, in the MCS model, the SC has two different mechanisms: along charge stripes for pairing and perpendicular to stripes for establishing the coherent state. As a consequence, carriers exhibit different properties in different directions: *fermionic* along charge stripes and *polaronic* perpendicular to stripes. It is important to note that charge stripes must fluctuate in order to avoid a charge-density-wave instability.[20,22] We found that many experimental data can be explained by the MCS model.[23] Moreover, the MCS model may explain the SC in an electron-doped cuprate.[3,22,23] However, it is possible that there is another model of HTSC which can explain experimental data better. Unfortunately, we are not aware of such model.

## 4. THE SYMMETRIES OF THE TWO GAPS

There is a consensus that the predominant OP in hole-doped cuprates has the d-wave symmetry.[1,7] Thus, one gap out of the two gaps has the d-wave symmetry. Since the $\Delta_c$ has most likely the magnetic origin this implies that it has the d-wave symmetry.[7] The second OP has to have partially or entirely a s-wave symmetry in order to explain in-plane torque anisotropy measurements.[24] More information on this issue can be found elsewhere.[25,26]

## 5. CONCLUSIONS

In summary, we discussed the phase diagram of order parameters for pairing and phase coherence in hole-doped cuprates. By examining some recent inelastic neutron scattering data obtained on hole-doped cuprates and on a heavy fermion compound $UPd_2Al_3$ in which the superconductivity is mediated by spin fluctuations, we concluded that the coherent gap in hole-doped cuprates has most likely the magnetic origin and scales with $T_c$, on average, as $2\Delta_c/k_B T_c = 5.4$. We discussed the MCS model of the superconductivity in hole-doped cuprates and the symmetries of the two order parameters.

# The Order Parameters in Cuprates

## ACKNOWLEDGMENTS

The author thanks R. Deltour for discussion. This work is supported by PAI 4/10 and FNRS.